\documentclass[astron]{iau} 
\usepackage{graphicx}
\usepackage{natbib}
\usepackage{amsmath}

\newcommand\au{{\,\mathrm{au}}}
\newcommand\pc{{\,\mathrm{pc}}}
\newcommand\msun{{\,\mathrm{M_\odot}}}
\newcommand\myr{{\,\mathrm{Myr}}}

\title[Three-body encounters in galactic nuclei] 
{Do three-body encounters in galactic nuclei affect compact binary merger rates?}

\author[Alessandro A. Trani]   
{Alessandro A. Trani$^1$
}

\affiliation{$^1$Department of Astronomy, Graduate School of Science, The University of Tokyo, \\7-3-1 Hongo, Bunkyo-ku, Tokyo, 113-0033, Japan \\ email: {\tt aatrani@gmail.com} \\[\affilskip]
}

\pubyear{2019}
\volume{351}  
\jname{Star Clusters: From the Milky Way to the Early Universe}
\editors{A. Bragaglia, M.B. Davies, A. Sills \& E. Vesperini, eds.}
\begin{document}

\maketitle

\begin{abstract}
	High-density cusps of compact remnants are expected to form around supermassive black holes (SMBHs) in galactic nuclei via dynamical friction and two-body relaxation. 
	Due to the high density, binaries in orbit around the SMBH can frequently undergo close encounters with compact remnants from the cusp. This can affect the gravitational wave merger rate of compact binaries in galactic nuclei.
	We investigated this process by means of high accuracy few-body simulations, performed with a novel Monte Carlo approach.
	We find that, around a SgrA*-like SMBH, three-body encounters increase the number of mergers by a factor of 3. This occurs because close encounters can reorient binaries with respect to their orbital plane around the SMBH, increasing the number of Kozai-Lidov induced mergers. We obtain a binary black hole merger rate of $\Gamma_\mathrm{MW} = 1.6 \times 10^{-6} \,\rm yr^{-1}$ per Milky Way-like nucleus.


	

\end{abstract}

\firstsection 
\section{Introduction}



Stellar binaries in orbit around supermassive black holes (SMBHs) will undergo secular oscillations in eccentricity and inclination known as Kozai-Lidov cycles. Several studies have shown that the excitation of eccentricity from the Kozai-Lidov mechanism will lead to an increased merger rate of compact binaries around SMBHs \citep[e.g.][]{anto12,hoa18,ham18}. 

However, previous studies have neglected the impact of close three-body encounters on the evolution of such binaries. In fact, a dense cusp of compact remnants is expected to grow around SMBHs via dynamical friction \citep[e.g.][]{alex09}. Recent observations suggest the existence of such a dense cusp in our Galactic center \citep{hail18}. Due to the high density, binaries in orbit around the SMBH can frequently undergo close three-body encounters with the black holes from the cusp, which will interrupt the secular Kozai-Lidov evolution of the binary.

Here we investigate the impact of three-body encounters on the merger rates of black hole binaries in orbit around a SMBH by means of hybrid N-body/Monte Carlo simulations.

\begin{figure}[bh]
\begin{center}

 \includegraphics[width=0.496\linewidth]{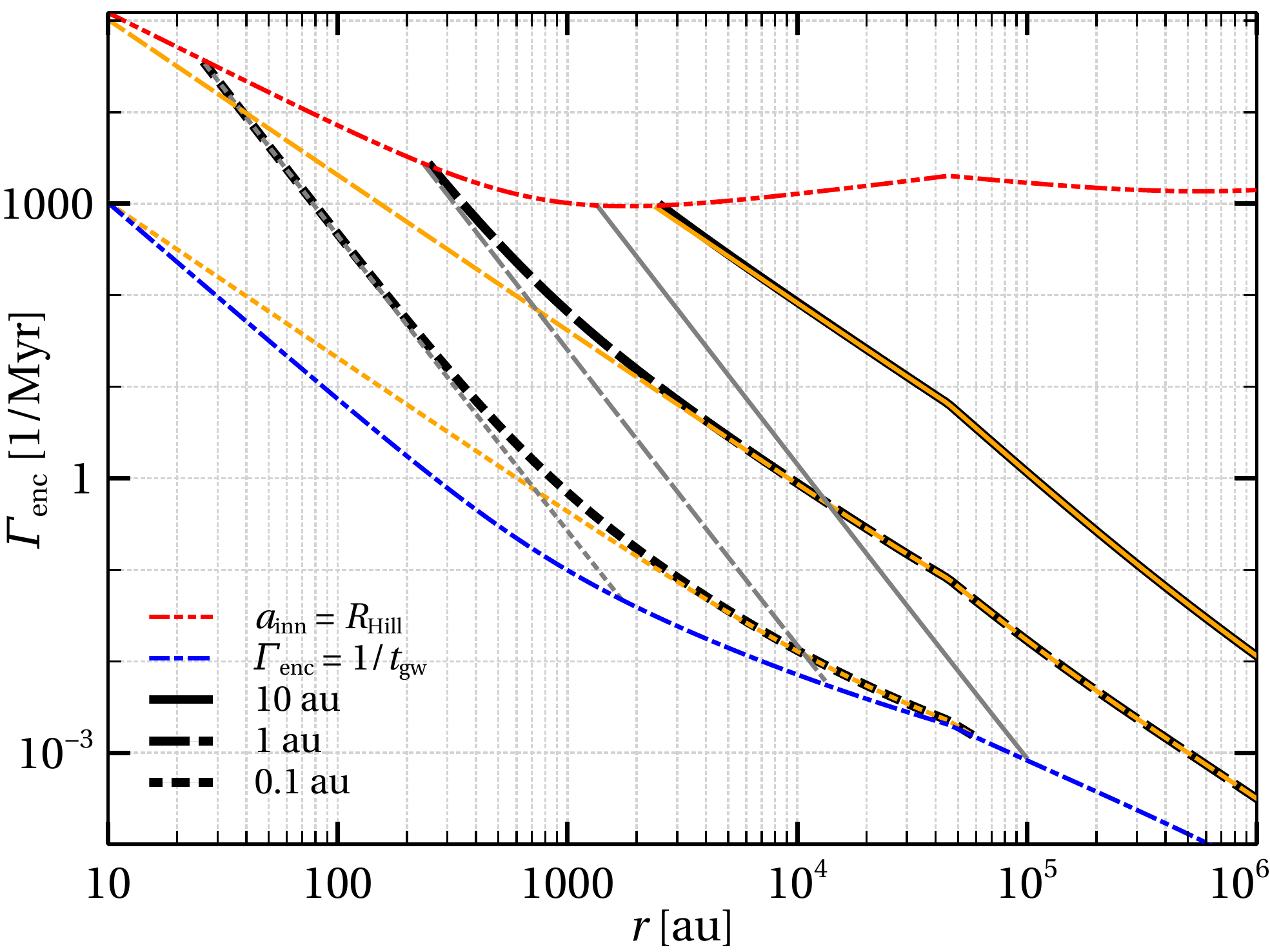}
 \includegraphics[width=0.496\linewidth]{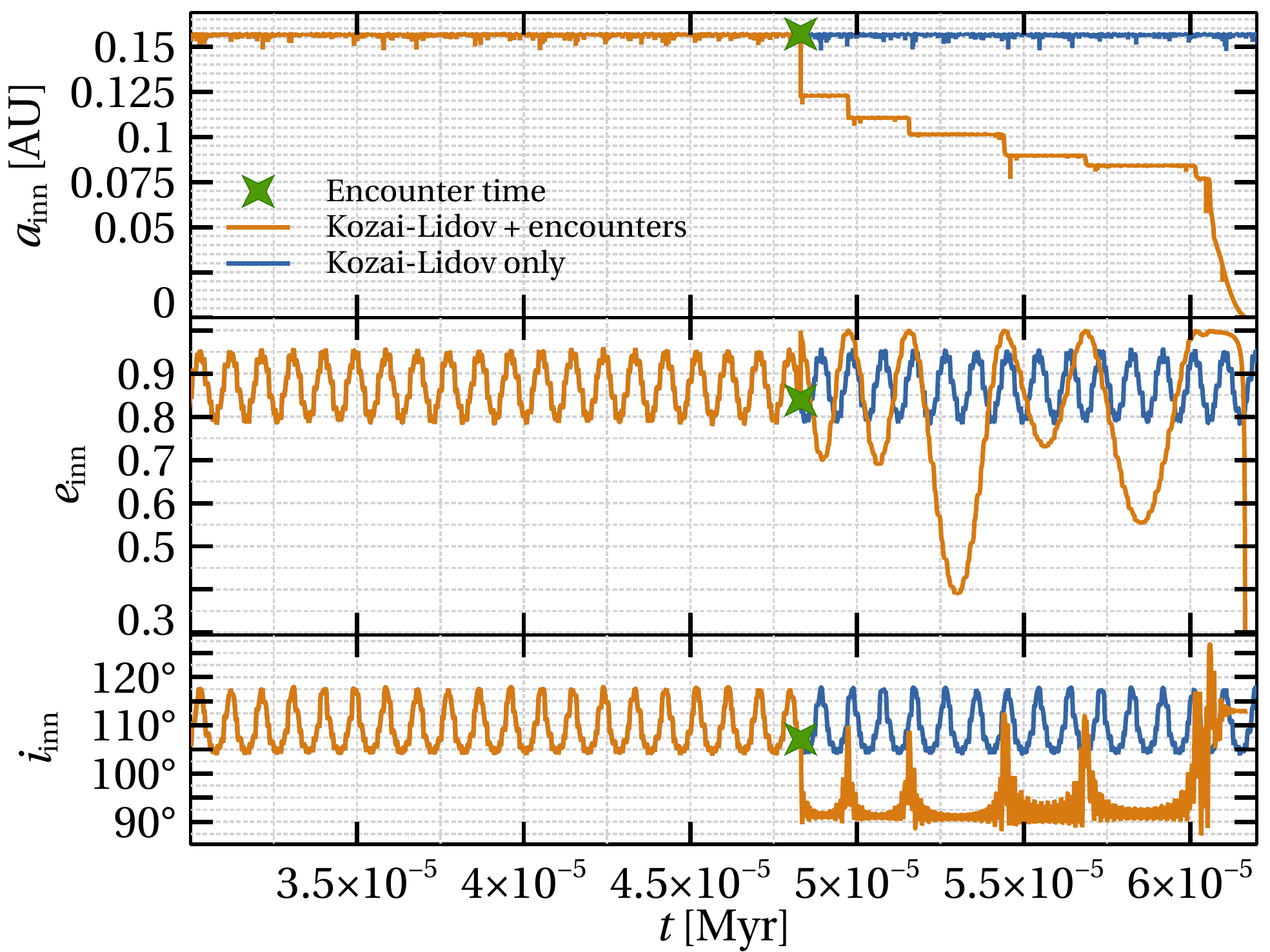} 
 \vspace*{-0.45 cm}
 \caption{
 	\footnotesize 
 	 	Left panel: encounter rates as a function of distance from the SMBH for binaries with semimajor axis $a_\mathrm{inn} = 10 \au$ (thick solid black line), $1 \au$ (thick dashed black line) and $0.1 \au$ (thick dotted black line). The yellow and grey lines are the contributions to the total encounter rate from the stars and from the compact remnants from the cusp, respectively. The dot-dashed red line is the upper limit to the encounter rate, set by the largest semimajor axis allowed by the tidal field (equivalent to the binary Hill radius). The dot-dashed blue line denotes the region below which the encounter timescale is longer than the merger time due to gravitational waves (assuming an equal mass, circular binary with total mass $60 \msun$).
 	Right panel: evolution of a binary's semimajor axis (top), eccentricity (middle) and inclination (bottom) as a function of time. Orange indicates the simulation including three-body encounters, while the blue line corresponds to the same simulation with no encounters. The green marker indicates the time of the encounter. 
 }
   \label{fig:encrateto}
\end{center}
\end{figure}

\section{Numerical setup}
\subsection{The \texttt{PROMENADE} code}
Running a direct gravitational N-body simulation of a nuclear star cluster with a central SMBH is not computationally feasible, due to the large number of particles and the high mass ratio between the SMBH and stars. Therefore, we have developed a new hybrid N-body/Monte Carlo code to incorporate the two main mechanisms that can trigger the gravitational wave coalescence of a compact binary in a galactic nucleus: (1) the Kozai-Lidov evolution due to the secular gravitational interactions with the SMBH and (2) three-body encounters with background stars and compact remnants from the nuclear star cluster.

Here we briefly describe our code, which we named \texttt{PROMENADE}. 
Each simulation follows the evolution an individual binary along its orbit around the SMBH. The binary-SMBH system is directly integrated with the fewbody code \texttt{TSUNAMI} (see \citealt{trani2019a,trani2019b} and \citealt{mik99a}), which includes post-Newtonian corrections up to the 2.5PN order. Initially, the binary is isolated in its motion around the SMBH, i.e. the binary-SMBH system forms an isolated hierarchical triple. After each timestep $\Delta t$ (typically of the order of the binary orbital period) we calculate the three-body encounter rate $\Gamma_\mathrm{enc}$ from the local density $n$ and velocity dispersion $\sigma$ of the assumed profile of the cusp \citep[see equation 9 of][]{lei16a}. With this we compute the probability for an encounter to occur as $P_\mathrm{enc}=\Gamma_\mathrm{enc}\, \Delta t$. Given the probability $P_\mathrm{enc}$, we can determine if an encounter will occur via Monte Carlo methods. If an encounter is determined to occur, we add to the simulation a fourth body in a interacting orbit with the binary, using the same prescriptions described in \citet{trani2019a}. The mass and orbit of the fourth body are determined from the assumed density profile of the nuclear star cluster. After the encounter is concluded and the fourth body is far from the binary, it is removed from the simulation and the isolated evolution is resumed.

\subsection{The initial conditions}
Here we present the results from a set of 1500 simulations of black hole binaries in a Galactic center-like environment. The SMBH mass is set to $4.31 \times 10^6 \msun$ \citep{gil17}. The masses of the two black holes are independently drawn form a log-uniform distribution between 6 and $150 \msun$. The semimajor axis $a_\mathrm{inn}$ of the binary is drawn from a log-uniform distribution between $0.1$ and $50 \au$, while the eccentricity $e_\mathrm{inn}$ is sampled from a uniform distribution. The binary inclination $i_\mathrm{inn}$ with respect to its orbital plane around the SMBH is isotropically distributed. 
The orbit of the binary around the SMBH has a semimajor axis $a_\mathrm{out}$ drawn from a power-law distribution with index $-1.93$ \citep{do13} and eccentricity $e_\mathrm{out}$ drawn from a normal distribution with $\mu = 0.3$ and $\sigma = 0.1$.

The background density profile is the sum of two components: the stellar cusp, which follows the broken power-law density profile from \citet{sch07}, and the cusp of compact remnants, which follows a power-law distribution with index $-11/4$ \citep{alex09} and has a total mass of $4\times 10^4 \msun$ between $50\au$ and $0.2\pc$, consistent with the constraints obtained from the orbits of the S-stars \citep{gil17}. 
For the Monte Carlo generated encounters, the semimajor axis distribution of the fourth body's orbit around the SMBH follows the same power law of the density profile, the eccentricity distribution is thermal and the inclination distribution isotropic. The mass function of the stellar component is a Salpeter distribution between $0.5$ and $2 \msun$, while the mass of the black holes from the cusp is uniformly sampled between $5$ and $50 \msun$.
The left panel of Figure~\ref{fig:encrateto} shows the encounter rates as function of the distance from the SMBH obtained with this setup.

We run the same set of initial conditions twice: once using the Monte Carlo approach for encounters (we refer to this set as KL+ENC), and a second time disabling the encounters (we refer to this set as KL). For the KL+ENC set, we integrate until either the binary merges, breaks up or a total integration time of 1 Myr is reached. For the KL set, we add another condition: since by disabling encounters we do not capture the binary breakup physics, we stop the integration when the total integration time reaches the binary evaporation timescale \citep[see equation 3 of][]{hoa18}. 

\section{Kozai-Lidov$\,+\,$encounters versus Kozai-Lidov only}
The right panel of Figure~\ref{fig:encrateto} compares the evolution of a binary's orbital parameters in a individual simulation with and without encounters. In this case, a single encounter triggers the coalescence of the binary in an extremely short time. In contrast, by neglecting the encounter, the binary would only merge in ${\approx}121 \myr$ via gravitational wave radiation only, a time much longer than its evaporation timescale.

\begin{table}[hb]
	\begin{center}
		\caption{\footnotesize Fractions of outcomes after $1 \myr$ of evolution}
		\label{tab:out}
		{\scriptsize
			\begin{tabular}{lcccccc}\hline 
				{\bf Set } & $f_\mathrm{break}^\mathrm{tot}$ & $f_\mathrm{break}^\mathrm{isol}$ & $f_\mathrm{break}^\mathrm{enc}$ & $f_\mathrm{merg}^\mathrm{tot}$ & $f_\mathrm{merg}^\mathrm{bin}$ & $f_\mathrm{merg}^\mathrm{pert}$ \\ \hline
				
				Kozai-Lidov + encounters & 0.637 & 0.418  & 0.219 & 0.311 & 0.304 & 0.007 \\
				Kozai-Lidov only         & 0.851 & 0.027 & 0.824$^1$ & 0.105 & 0.105 & -- \\  \hline
			\end{tabular}
		}
	\end{center}
	\vspace{1mm}
	\scriptsize{
		$f_\mathrm{break}^\mathrm{tot}$: total fraction of breakups; $f_\mathrm{break}^\mathrm{isol}$: fraction of breakups occurred in isolation; $f_\mathrm{break}^\mathrm{enc}$: fraction of breakups during encounters; $f_\mathrm{merg}^\mathrm{tot}$: total fraction of black-hole binary mergers; $f_\mathrm{merg}^\mathrm{bin}$: fraction of mergers involving the original binary; $f_\mathrm{merg}^\mathrm{pert}$: fraction of mergers between one member of the original binary and the perturber.\\
		{$^1$: using the evaporation timescale criterion.} 
	}
\end{table}

In Table~\ref{tab:out} we summarize the outcomes of the simulations. Allowing for encounters almost triples the number of mergers within the same timeframe: about 31\% mergers occur in the KL+ENC set, and only about 10\% occur in the KL set. Another striking difference is the number of binary breakups: the using the simple evaporation timescale criterion leads to about one-third more binary breakups than when fully modelling the encounters. 

It is worth noting that the majority of the mergers still occurs during the isolated Kozai-Lidov evolution: only about $2.2\%$ of the mergers occur during an encounter, and only between a black hole from the binary and the interloping one. We can explain this by considering how the encounters can trigger the gravitational wave coalescence: (i) either by increasing the binary eccentricity and shrinking the binary semimajor axis, thus decreasing the gravitational radiation emission timescale, or (ii) by altering the orientation of the binary so that the Kozai-Lidov mechanism can take place. The ratio $f_\mathrm{break}^\mathrm{enc}/f_\mathrm{break}^\mathrm{isol}$ gives us a hint to which of these two processes is more efficient. Since the majority of the binaries breaks up only long after an encounter has taken place, we believe that the encounters are triggering Kozai-Lidov-induced mergers, rather than hardening the binaries until coalescence (as it occurs in star clusters). This will be thoroughly demonstrated in our coming work.

\section{Conclusions}
The simple models presented here illustrate the impact of encounters on the Kozai-Lidov evolution of compact binaries in orbit around SMBHs, an aspect that was marginally considered in previous studies. 
Three-body encounters greatly enhance the binary black hole merger rate in galactic nuclei. In our Milky Way-like setup, including the encounters increases the merger fraction by a factor of 3 compared to when neglecting the encounters. 
We attribute the increase in mergers to the encounters widening the parameter space for which Kozai-Lidov-induced mergers take place, a process we may term ``encounter-assisted Kozai-Lidov merger''.

Moreover, by modeling the three-body encounters we capture the correct binary breakup physics and obtain a more accurate number of binary breakups than by using the simple evaporation timescale argument.

Assuming a black hole formation rate of $\Gamma_\mathrm{BH} = 10^{-4}\,\rm yr$, a black hole binary fraction of $f_\mathrm{BH} = 0.05$, we obtain a merger rate of $\Gamma_\mathrm{MW} = \Gamma_\mathrm{BH}\,f_\mathrm{BH}\,f_\mathrm{merg}^\mathrm{tot} = 1.6 \times 10^{-6} \,\rm yr^{-1}$ per Milky Way-like nucleus. In our coming work, we will present our complete set of simulations, considering different SMBH masses, realistic binary distributions and different cusp profiles.

\bibliographystyle{astron}
\bibliography{ms}


\end{document}